\documentclass[11pt,a4paper]{article}

\usepackage[utf8]{inputenc}
\usepackage[T1]{fontenc}
\usepackage{lmodern}
\usepackage[margin=2.3cm]{geometry}
\usepackage{graphicx}
\usepackage{amsmath,amssymb}
\usepackage{booktabs}
\usepackage{array}
\usepackage{caption}
\usepackage{microtype}
\usepackage{authblk}

\usepackage{xcolor}
\usepackage{placeins}
\usepackage[bookmarks=false]{hyperref}
\hypersetup{hidelinks}

\title{\bfseries Articulate but Wrong:\\
Self-Review Failures in LLM-Based Code Modernization}

\author{Gokul Chandra Purnachandra Reddy\thanks{Corresponding
author: \texttt{gokulcpr@ieee.org}. The views expressed are the
authors' and do not necessarily reflect Amazon Web Services or
Amazon.com, Inc. This research was conducted independently of the
authors' employment and did not use any proprietary or employer-
confidential data, systems, or resources.}}
\author{Aditya Lolla}
\author{Harsha Sanku}
\affil{Amazon Web Services, California, USA}

\date{May 2026}

\begin{document}
\maketitle

\begin{abstract}
Large language model (LLM) agents are increasingly used to migrate
legacy code to modern stacks. We ask a deceptively simple question:
when an LLM modernizes legacy code, can the same model be relied upon
to recognize when its own output silently changes observable
behavior? We run 1{,}980 real modernization calls across 11 production
LLMs from 7 distinct families on a balanced 60-snippet legacy-Python-2
corpus, evaluate every output with a type-strict behavioral oracle,
and then ask each model to judge whether its own output preserves
behavior. We report four findings. (1) Semantic-preservation drift is
prevalent and sharply separable from a cleanly-controlled baseline:
semantic-trap snippets drift in 39.7\% of attempts versus 7.0\% on
benign-control code that requires no real modernization
($+32.7$\,percentage points; $n=660$ each). (2) Drift concentrates on
specific snippets that fail across models: pairwise model agreement on
which snippets are hard is high (mean Pearson $r=0.52$), and a small
core of numeric-semantics snippets fails for nearly every model and
every prompt phrasing. (3) Self-review by the producing model is not
a reliable safety net: across all semantic drift cases, 31.7\% are
silently endorsed by the same model that produced them ($83/262$),
and the per-model self-miss rate is strongly bimodal -- ranging from
$0\%$ on five models to $100\%$ on one widely deployed model -- with
several models explicitly articulating the very Py2/Py3 semantic
distinction that broke their output, then declaring behavior
preserved. (4) Drift rate is non-monotone in model capability and
price: per-model rates range $5.6\%$--$46.7\%$ and do not track model
capability cleanly, indicating the failure is task-structural rather
than driven by model scale. All code, prompts, the 60-snippet corpus,
the behavioral oracle, the output extractor, and the raw model
outputs are released.
\end{abstract}


\section{Introduction}
\label{sec:intro}
Multi-agent LLM pipelines now perform repository-scale code
modernization at substantial pass rates~\cite{legacytranslate,vapu,
envloop}. Across that literature, however, one limitation recurs:
generated migrations come with no correctness assurance, and
establishing one still requires expensive human review~\cite{mitremod}.
The most natural-feeling cheap assurance -- ask the same model
whether its output preserves behavior -- is exactly what we test
here, on the most insidious failure mode: a migration that compiles,
runs without exception, and produces \emph{plausible but different}
outputs.

\paragraph{Why the question matters.}
Code-modernization failures fall on a spectrum. At one end,
syntactic mistakes (a removed API, a renamed module) are caught
trivially because the code does not run. At the other end, silent
behavioral drift -- the modernized code runs, returns values, and
quietly disagrees with the legacy contract on some inputs -- is the
class operators most fear, because no compile/test-pass check catches
it. Whether the producing LLM itself recognizes such drift, when
asked, is the question that determines whether self-review is even
plausible as a safety net.

\paragraph{What we do.}
We construct a balanced 60-snippet legacy-Python-2 corpus across three
trap categories: semantic-preservation traps (numeric semantics, lazy
evaluation, type model), syntactic traps (API removal, literal
syntax), and a benign-control category requiring \emph{no real
modernization} (e.g.\ \texttt{return a+b}). We run 1{,}980 real
modernization calls across 11 LLMs from 7 distinct families and 3
prompt phrasings, evaluate each output with a type-strict behavioral
oracle in the tradition of testing-equivalence semantics
\cite{denicola1984}, then ask the producing model to judge whether
its own output preserved behavior. We analyse drift incidence,
per-class structure, cross-model agreement on which snippets are
hard, prompt sensitivity, the role of model capability, and the
sensitivity, specificity, and per-model variation of self-review.

\paragraph{Contributions.}
\begin{itemize}
\item A reproducible benchmark with a cleanly-controlled benign
baseline, a type-strict behavioral-equivalence oracle that catches
the int $\to$ float drift that \texttt{==} would hide, and 11
production LLMs spanning a wide capability and cost range
(Section~\ref{sec:method}).
\item Evidence that semantic-preservation traps drift at 39.7\%
versus a benign-control baseline of 7.0\% ($+32.7$ pp,
$n=660$ each), with numeric-semantics traps dominating at 57\%
(Section~\ref{sec:drift}).
\item Evidence that drift is task-structural: pairwise cross-model
agreement on which snippets are hard has mean Pearson $r=0.52$, and
a small core of snippets fails for nearly every model and prompt
(Section~\ref{sec:drift}).
\item The paper's headline result: self-review by the producing
model misses 31.7\% of its own semantic drift, with strongly bimodal
per-model behaviour (0\%--100\%) including verbatim cases in which
models explicitly articulate the Py2/Py3 difference and then declare
behaviour preserved (Section~\ref{sec:self}).
\item Two methodological lessons: drift rate is non-monotone in model
capability and price (Section~\ref{sec:capability}), and the choice
of behavioural oracle and output extractor each materially affects
the measured drift rate; we document these counterfactuals
(Section~\ref{sec:method-comparison}).
\end{itemize}

\section{Related Work}
\label{sec:related}
LLM-based code translation and modernization is now an active
applied area~\cite{legacytranslate,vapu,envloop}, with most evaluation
in aggregate compile-and-test-pass terms. A recent study identifies
absence of automated correctness assurance as the principal
practical barrier~\cite{mitremod}. A complementary line proposes
\emph{verifying} agent outputs against external policies or formal
specifications~\cite{veriguard,formaljudge}.

Our notion of preservation is observational, in the tradition of
testing equivalence and behavioural semantics: two programs are
equivalent if no observable test distinguishes them
\cite{denicola1984}. We make the equivalence concrete with a
type-strict input/output oracle rather than a process-algebraic
relation, because the question we answer is about the
\emph{behaviour visible to a downstream caller}, not about full
program equivalence. Our approach is also adjacent to
\emph{mutation testing}, which studies the ability of test suites to
detect small behaviour-changing modifications~\cite{jiaharman2011};
silent drift produced by LLM modernization can be viewed as a
naturally-occurring mutation that the developer did not intend.

The self-review probe we introduce is, in spirit, the simplest
member of a family of self-consistency and self-evaluation methods
for LLMs. Self-consistency over multiple sampled chains-of-thought
has been shown to improve reasoning accuracy in arithmetic and
commonsense tasks~\cite{wang2023sc}. Whether self-review provides a
useful safety net in code modernization -- where the property to
preserve is behavioural rather than answer-level -- is, to our
knowledge, an open question, and the population-level and bimodal
patterns we report have not been documented in this setting.

\section{Method}
\label{sec:method}

\paragraph{Corpus.}
We construct $60$ legacy Python-2 snippets in three balanced groups
of $20$:
\begin{itemize}
\item \textbf{Semantic} ($n=20$): legacy idioms whose modernization
can silently change observable behaviour. Three sub-classes: numeric
semantics ($/$, integer mean, century rounding), lazy-eval
(\texttt{map}, \texttt{filter}, \texttt{zip}, \texttt{range},
\texttt{dict.keys/items} as indexable lists), and type model
(\texttt{long} vs \texttt{int}, \texttt{str}/\texttt{unicode}).
\item \textbf{Syntactic} ($n=20$): legacy idioms whose Py3 equivalent
is unambiguous (\texttt{has\_key}, \texttt{xrange}, \texttt{reduce}
moved to \texttt{functools}, \texttt{cmp=} sort, octal/unicode
literals, etc.).
\item \textbf{Benign-control} ($n=20$): code that does \emph{not}
require semantic or API modernization and runs identically under
Py2 and Py3 (\texttt{a+b}, \texttt{max}, \texttt{sum},
\texttt{str.upper}, etc.). This category cleanly measures the
baseline rate at which LLMs introduce drift even when no
modernization is actually warranted.
\end{itemize}
Every snippet defines a function \texttt{solve} with a fixed
input/output contract validated against the Python-2 semantics it
encodes; the corpus is released.

\paragraph{Behavioral oracle.}
A modernized candidate is evaluated against the legacy contract using
\emph{type-strict equality}: same type and same value (recursively
through containers). Strict equality is necessary because the most
prevalent drift is \texttt{int}~$\to$~\texttt{float}
(e.g.\ Py2 \texttt{5/2}~$=$~\texttt{2} versus Py3
\texttt{5/2}~$=$~\texttt{2.5}); under permissive \texttt{==}, several
of these would falsely register as preserved because \texttt{2 == 2.0}
in Python's value-based comparison. Section~\ref{sec:method-comparison}
documents the magnitude of this effect quantitatively. The oracle is
the empirical analogue of an observational test
\cite{denicola1984}: behaviour is what the caller can see, including
the type tag of the returned value.

\paragraph{Models.}
We use 11 production LLMs accessed through OpenRouter, spanning seven
distinct model families and a $\sim$200$\times$ cost range
(Table~\ref{tab:models}). Models are referenced in results by
anonymous tags M1--M11 chosen to keep the focus on the population-
level phenomenon rather than vendor comparison; the mapping is
provided for reproducibility but the analysis does not depend on it.

\begin{table}[!htbp]
\centering
\small
\begin{tabular}{lll}
\toprule
Model & Family & Output price (USD / 1M tokens) \\
\midrule
gpt-5.1 & OpenAI & 10.00 \\
gemini-2.5-pro & Google & 10.00 \\
gemini-2.0-flash-001 & Google & 0.40 \\
claude-3-haiku & Anthropic & 1.25 \\
deepseek-chat-v3.1 & DeepSeek & 0.79 \\
llama-3.3-70b-instruct & Meta & 0.32 \\
llama-3.1-8b-instruct & Meta & 0.05 \\
qwen-2.5-72b-instruct & Alibaba & 0.40 \\
qwen-2.5-coder-32b-instruct & Alibaba & 1.00 \\
mistral-small-3.2-24b-instruct & Mistral & 0.20 \\
command-r-08-2024 & Cohere & 0.60 \\
\bottomrule
\end{tabular}
\caption{Production LLMs used. Pricing from OpenRouter at run time.
Results are reported as M1--M11 (anonymized by sorted model id) to
keep focus on the phenomenon rather than ranking.}
\label{tab:models}
\end{table}

\paragraph{Prompts.}
We test three distinct phrasings of the same task: \texttt{P1\_direct}
(a direct behaviour-preserving instruction), \texttt{P2\_minimal}
(``port this; keep behavior identical; output only code''), and
\texttt{P3\_contract} (a contract-framed production instruction
naming downstream callers). Prompts are released. Each snippet is run
under each prompt by each model: $20 \times 3 \times 11 = 660$ calls
per trap category, $1{,}980$ total. Temperature is $0$;
\texttt{max\_tokens} is $700$.

\paragraph{Output extraction.}
Model responses are post-processed by a robust extractor that
identifies the modernized \texttt{solve} function regardless of
markdown fences, leading explanatory prose (``Here's the Python 3
version:''), or trailing commentary. We document and release the
extractor; using a naive fence-only extractor inflates the apparent
benign-control drift rate from $7.0\%$ to roughly $23\%$ by
mis-attributing parsing failures as behavioural drift
(Section~\ref{sec:method-comparison}).

\paragraph{Self-review probe.}
After all modernizations are evaluated, we issue a separate judgement
call to the \emph{same model} that produced each candidate, supplying
the legacy snippet and the candidate code and asking for a YES/NO
judgement on behavioural preservation. The judge does not know the
candidate is its own prior output, mirroring the realistic
deployment scenario where a verification step runs after generation.

\begin{figure}[!htbp]
\centering
\includegraphics[width=\textwidth]{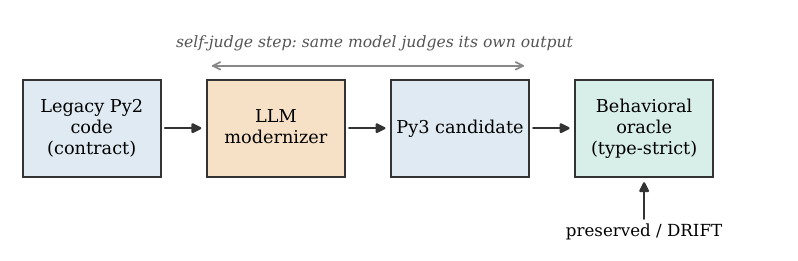}
\caption{The experimental setup. Each legacy snippet is modernized by
each model under each prompt, evaluated by a type-strict behavioural
oracle against the legacy contract, and then the same model is asked
whether its own output preserved behaviour.}
\label{fig:concept}
\end{figure}

\section{Drift Structure}
\label{sec:drift}

\begin{figure}[!htbp]
\centering
\includegraphics[width=0.55\textwidth]{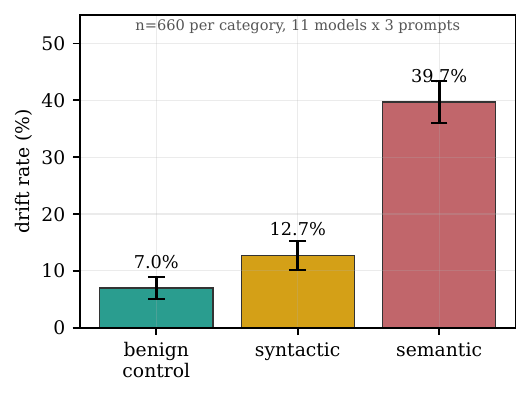}
\caption{Drift rate by trap category. Benign-control snippets require
no real modernization and measure the baseline rate at which LLMs
introduce silent behaviour change; semantic-preservation traps drift
at roughly $5.6\times$ the syntactic-excess rate.}
\label{fig:tiers}
\end{figure}

\paragraph{Three tiers with a cleanly-controlled baseline.}
Figure~\ref{fig:tiers} shows drift rates aggregated over all $660$
calls within each category. Benign-control drift is $7.0\%$
(95\% CI $\approx \pm 1.9$ pp), syntactic drift is $12.7\%$
($\pm 2.5$ pp), and semantic drift is $39.7\%$ ($\pm 3.7$ pp). The
syntactic excess over baseline is $+5.8$ pp; the semantic excess is
$+32.7$ pp -- semantic-preservation traps drift roughly five and a
half times more above baseline than syntactic traps do.

\begin{figure}[!htbp]
\centering
\includegraphics[width=0.62\textwidth]{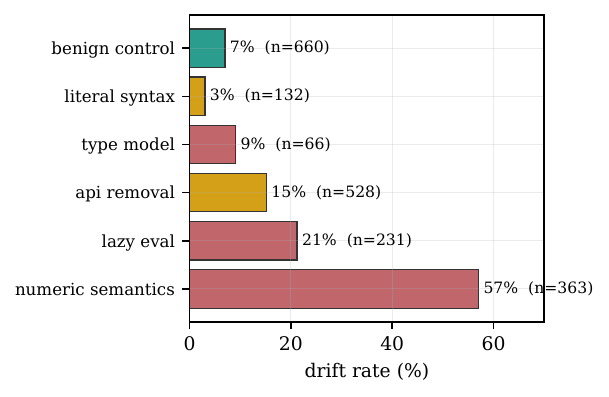}
\caption{Drift rate by specific trap class. Numeric-semantics
preservation is the dominant failure mode.}
\label{fig:perclass}
\end{figure}

\paragraph{Numeric semantics dominate.}
Figure~\ref{fig:perclass} breaks the rate down by trap class.
Numeric-semantics drift (Py2 \texttt{/}-as-floor on integer operands)
is the most prevalent failure mode at $57\%$, followed by lazy-eval
drift at $21\%$. Type-model and literal-syntax traps drift near
baseline.

\begin{figure}[!htbp]
\centering
\includegraphics[width=0.8\textwidth]{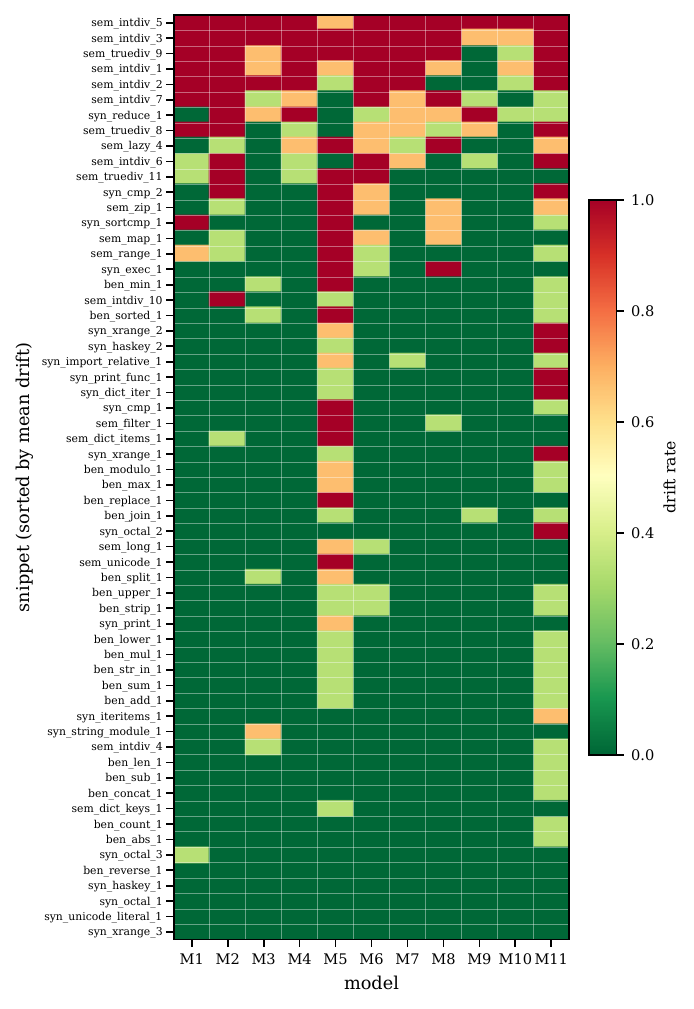}
\caption{Drift rate per snippet (rows, sorted by mean drift across
models) and per model (columns). The persistent core of numeric-
semantics snippets at the top is red across most models; benign and
trivial-syntactic snippets at the bottom are uniformly green.}
\label{fig:heatmap}
\end{figure}

\paragraph{Drift is task-structural, not random.}
Models do not drift on random subsets: the pairwise correlation
between models' per-snippet drift rates has mean Pearson $r=0.52$
($n=55$ pairs; median $r=0.57$; max $r=0.94$). The hardest snippets
are hard for almost every model (Figure~\ref{fig:heatmap}); the
easiest are easy for almost every model. A small core of numeric-
semantics snippets fails for $\ge 8/11$ models even under the most
favourable prompt phrasing. The shared structure of failure points to
the modernization \emph{task} -- specifically the Py2/Py3 division
semantics gap -- rather than to idiosyncratic model errors.

\paragraph{Prompt sensitivity sits at the margin, not the core.}
The three prompt phrasings produce different aggregate drift rates
but they do not eliminate the persistent core. Overall drift rates
under P1\_direct, P2\_minimal, and P3\_contract are
$\approx 19\%$, $\approx 24\%$, and $\approx 32\%$ respectively;
prompt P3, which emphasises a downstream-caller contract, paradoxically
yields the highest drift, consistent with the model producing more
elaborate ``safer'' rewrites that change semantics. Crucially the
persistent-core snippets drift for $\ge 8/11$ models \emph{under
every prompt}: prompt phrasing modulates the periphery but cannot
eliminate the structural failure on the hardest semantic traps.

\section{Self-Review is Not a Safety Net}
\label{sec:self}

For every modernization, we asked the same model to judge whether the
candidate preserved behaviour. We focus on the population of cases
where the oracle identified actual semantic drift ($262$ cases) and
ask: how often does the producing model agree with the oracle that
its own output drifted? The opposite outcome -- the model declares
behaviour preserved when in fact it is not -- is the dangerous one,
because it is precisely the outcome a self-review pipeline would
silently approve.

\begin{figure}[!htbp]
\centering
\includegraphics[width=0.6\textwidth]{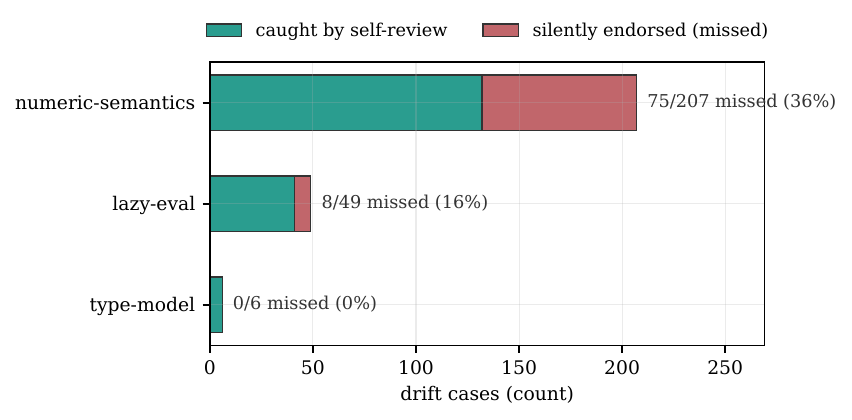}
\caption{Self-review outcomes on the model's own semantic drift,
broken down by trap class. Red indicates drifts the producing model
silently endorsed as behaviour-preserving.}
\label{fig:miss}
\end{figure}

\paragraph{Overall: $31.7\%$ silently endorsed.}
Across all $262$ semantic drift cases, $83$ ($31.7\%$) are silently
endorsed by the same model that produced them. Numeric-semantics
drift accounts for the bulk: $75$ of $207$ ($36\%$)
of these errors pass self-review (Figure~\ref{fig:miss}). The
lazy-eval class fares somewhat better ($8/49$, $16\%$ missed);
type-model is fully caught ($0/6$) but with very low statistical
weight.

\begin{figure}[!htbp]
\centering
\includegraphics[width=0.6\textwidth]{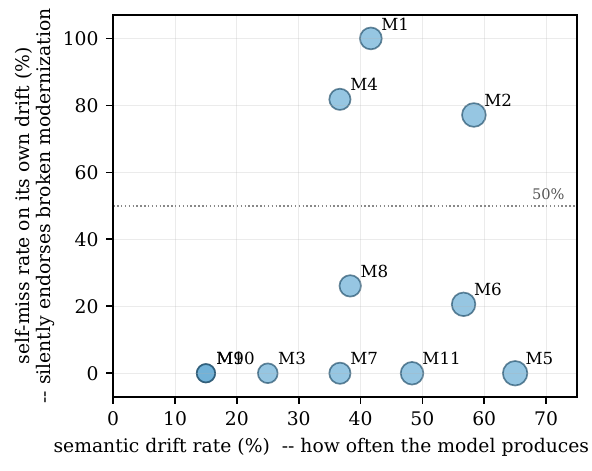}
\caption{Per-model semantic drift rate (x-axis) against the rate at
which the same model silently endorses its own drift (y-axis). The
self-miss rate is strongly bimodal and orthogonal to the drift rate.
M1, the model with $100\%$ self-miss, drifts only moderately; M5,
the highest-drifting model, catches all of its own drift.}
\label{fig:scatter}
\end{figure}

\paragraph{The per-model picture is bimodal.}
Aggregate rates conceal the most striking finding.
Figure~\ref{fig:scatter} plots, for each model, its semantic drift
rate ($x$) against the fraction of its own drift it then silently
endorses ($y$). Two facts are visible at a glance. First, the
self-miss rate is bimodal: five models miss essentially $0\%$ of
their own semantic drift; five miss $20$\%--$100\%$. Second, the
self-miss rate is not predicted by the drift rate -- M5 has the
highest drift ($65\%$) but a $0\%$ self-miss rate, while M1 has
moderate drift ($42\%$) and a $100\%$ self-miss rate. Self-review
reliability is therefore a per-model property orthogonal to
modernization correctness.

\subsection{Qualitative: articulating the gap, then declaring preservation}
\label{sec:qualitative}
The qualitative pattern on high self-miss models is the genuinely
surprising one. For numeric-semantics drift, several models open
their self-review by correctly stating the Py2/Py3 distinction --
and then continue with a \textsc{yes} declaration that behaviour is
preserved. The model articulates the exact mechanism that broke its
output, and concludes the output is unbroken.

\paragraph{Three verbatim examples.}
All three are from model M1 (anonymized; mapping in
Table~\ref{tab:models}) on numeric-semantics traps it actually
drifted on. Each judgement opens with \textsc{yes} (``behaviour
preserved'') and continues:
\begin{quote}\small\itshape
\noindent[sem\_intdiv\_1] ``The observable input/output behavior of
the legacy code is preserved exactly in the candidate modernized
version. In Python 2, the division operator \texttt{/} performs
integer division if both operands are integers, and floating-point
division if at least one operand is a float. This behavior is the
same in Python 3, so the modernized version will produce the same
results as the [legacy] \ldots''
\end{quote}
\begin{quote}\small\itshape
\noindent[sem\_intdiv\_2] ``The observable input/output behavior of
the legacy code is preserved exactly in the candidate modernized
version. In Python 2, the division operator \texttt{/} performs
integer division when both operands are integers, while in Python 3,
the same operator performs floating-point division. Since the
candidate code uses the same division operator, the behavior remains
the same \ldots''
\end{quote}
\begin{quote}\small\itshape
\noindent[sem\_intdiv\_3] ``The observable input/output behavior of
the legacy code is preserved exactly in the candidate modernized
version. Both the legacy code and the candidate code perform the
same mathematical operation \ldots The division operation in Python
3 behaves the same way as in Python 2, so the final result will be
\ldots''
\end{quote}

\noindent The first example correctly states the Py2/Py3 rule; the
second correctly states that the two languages \emph{differ}; the
third asserts they behave \emph{the same}. All three conclude
preservation. The pattern is consistent across high self-miss models
on the numeric-semantics class.

\paragraph{What this implies.}
Self-review by the producing model is not a usable safety net for
silent semantic drift in LLM code modernization, even though it
catches $\sim 68\%$ of semantic drift overall. The failure mode is
not low recall (which would be merely incomplete); it is silent
high-confidence endorsement on a substantial sub-population, and it
is concentrated on the most prevalent and consequential trap class.

\section{Drift Is Not Monotone in Model Capability}
\label{sec:capability}
A reasonable prior is that better-resourced models drift less. The
data does not support this expectation. Per-model semantic drift
rates range from $5.6\%$ to $46.7\%$, but they do not order by model
price or by family-internal capability tier. The lowest-priced model
in our panel (output \$0.05/1M tokens) drifts at $5.6\%$, lower than
several mid- and frontier-tier models, while one of the
highest-priced models in the panel drifts at $22.8\%$. Two
inferences follow. First, the dominant failure mode --
numeric-semantics drift -- is not a capability bottleneck that
scaling appears to fix in our snapshot. Second, vendor- or
size-based selection rules do not protect against this class of
drift; an explicit behavioural oracle is required. Per-model rates
are in Table~\ref{tab:per-model}.

\begin{table}[!htbp]
\centering
\small
\begin{tabular}{lrrrr}
\toprule
Tag & sem.~drift & benign drift & sem.\,excess (pp) &
self-miss on sem. \\
\midrule
M1  & 41.7\% &  0.0\% & $+41.7$ & 100.0\% \\
M2  & 58.3\% &  0.0\% & $+58.3$ & 77.1\% \\
M3  & 25.0\% &  5.0\% & $+20.0$ &  0.0\% \\
M4  & 36.7\% &  0.0\% & $+36.7$ & 81.8\% \\
M5  & 65.0\% & 38.3\% & $+26.7$ &  0.0\% \\
M6  & 56.7\% &  3.3\% & $+53.3$ & 20.6\% \\
M7  & 36.7\% &  0.0\% & $+36.7$ &  0.0\% \\
M8  & 38.3\% &  0.0\% & $+38.3$ & 26.1\% \\
M9  & 15.0\% &  1.7\% & $+13.3$ &  0.0\% \\
M10 & 15.0\% &  0.0\% & $+15.0$ &  0.0\% \\
M11 & 48.3\% & 28.3\% & $+20.0$ &  0.0\% \\
\bottomrule
\end{tabular}
\caption{Per-model semantic and benign drift, semantic excess over
each model's own benign baseline, and self-miss rate on the model's
own semantic drift. The ordering by sem.~excess does not match price
or capability tier (Table~\ref{tab:models}).}
\label{tab:per-model}
\end{table}

\section{Methodological Choices Materially Change the Result}
\label{sec:method-comparison}
Two methodological choices in our pipeline are not aesthetic; they
materially change the measured drift rate, and prior work that does
not control for them likely under-reports silent semantic drift.

\paragraph{Permissive equality undercounts numeric drift.}
Repeating the analysis with permissive \texttt{==} comparison instead
of type-strict equality lowers the measured semantic drift from
$39.7\%$ to approximately $26.7\%$, a $\sim$13 percentage point
reduction. The lost cases are exactly those where Py2 \texttt{5/2 = 2}
becomes Py3 \texttt{2.5} and \texttt{2 == 2.0} evaluates as
\texttt{True}. A downstream consumer that does
\texttt{isinstance(x, int)}, that uses the value as a list index, that
serialises it as JSON, or that compares it strictly will see
different behaviour; the permissive \texttt{==} hides this. We
therefore use strict equality, in line with the observational view of
behaviour: the type tag is observable~\cite{denicola1984}.

\paragraph{Fence-only output parsing inflates the apparent baseline.}
Many models, especially when prompted with the production-flavoured
P3\_contract, return modernized code wrapped in explanatory prose
(``Here is the Python 3 version of the function: \ldots'').
A fence-only parser that extracts \texttt{```python\ldots```} blocks
or otherwise treats raw model output as the candidate code fails to
parse a substantial fraction of these responses; the harness then
mis-attributes the parse failure as a behavioural drift. With a
fence-only parser, the benign-control drift rate is approximately
$23\%$; with the robust extractor we release, it falls to $7.0\%$.
The fence-only number is what a naive pipeline would observe and is
the figure several prior reports would have measured had they used a
behavioural oracle. The discrepancy is methodological, not
behavioural, and the robust extractor is therefore part of the
published artifact rather than a convenience.

\FloatBarrier
\section{Discussion and Limitations}
\label{sec:limits}

\paragraph{What the result says, and what it does not.}
The result is negative and behavioural. It does not say LLMs cannot
modernize code -- many migrations succeed -- and it does not rank
models. It says that, for the most prevalent silent-drift class,
asking the producing model to verify itself is not a substitute for
behavioural testing against the legacy contract, that drift is
task-structural rather than model-monotone, and that two
methodological choices materially affect what one measures.

\paragraph{Limitations (stated plainly).}
(i) Our corpus is in Python and focused on the Py2$\to$Py3 transition,
the cleanest legacy/modern boundary with mechanically checkable
contracts. Generalising to other language transitions is open.
(ii) Snippets are short ($\le 10$ lines), which understates the
combinatorial complexity of real modernization; we expect the drift
\emph{class} structure to transfer but not the absolute rates.
(iii) We test only single-model self-review, not multi-model voting,
chain-of-thought-augmented review~\cite{wang2023sc}, or tool-augmented
review -- those remain open questions, and the present finding is
specifically a floor: \emph{plain} self-review is not sufficient.
(iv) Model identities behind M1--M11 are released for reproducibility
(Table~\ref{tab:models}) but the analysis is deliberately framed at
the population level. (v) Behavioural equivalence is a stronger
notion than test-pass~\cite{denicola1984,jiaharman2011}; we report it
as the more conservative criterion, and observe that several drifts
would pass naive \texttt{==} comparison.

\paragraph{Use of generative AI.}
In accordance with COPE guidance, the authors disclose that
generative-AI tools were used for mechanical and editorial
assistance only: code commenting, language editing, and \LaTeX{}
formatting. The problem formulation, the experimental design, the
behavioural oracle, the analysis, and the interpretation are the
authors' own work. Generative AI was not used to generate data,
results, or scientific conclusions, and is not credited as an
author. The authors take full responsibility for the integrity and
accuracy of the work.

\section{Conclusion}
LLM code modernization exhibits a high baseline of silent
behavioural drift on semantic-preservation traps -- particularly
numeric-semantics traps -- that is task-structural rather than
random and that does not track model capability cleanly. Self-review
by the producing model catches most drift in aggregate but silently
endorses approximately one in three semantic drifts, with bimodal
per-model behaviour: several models confidently approve their own
broken modernizations, sometimes while correctly articulating the
exact semantic distinction that broke them. Two methodological
choices -- a type-strict behavioural oracle and a robust output
extractor -- materially change the measured rate and should be
treated as part of the experimental design. We therefore caution
against deploying single-model self-review as a safety net for
silent semantic drift in production code-modernization pipelines.

\section*{Data and Code Availability}
The complete reproducibility package -- the 60-snippet
legacy-Python-2 corpus with its Py2-semantic contracts
(\texttt{dataset\_v2.py}), the type-strict behavioural oracle
(\texttt{oracle.py}), the robust output extractor
(\texttt{extract\_code.py}), the parallel modernization runner
(\texttt{harness\_v2p.py}), the self-review probe
(\texttt{self\_detect.py}), the three prompt phrasings and the
11-model panel specification (\texttt{robust.py}), and the raw
JSON-lines logs of all $1{,}980$ modernization calls and $1{,}979$
self-review calls -- is publicly archived at
\url{https://zenodo.org/records/20300861}, together
with a \texttt{README.md} giving the exact command sequence to
reproduce every figure and table in this paper. Using the per-call
token-usage logs included in the JSON-lines files, the total
OpenRouter API cost to re-run the full experiment is approximately
\textsc{usd}~3.40 at the prices we logged at run time.

\section*{Conflicts of Interest}
The authors are employed by Amazon Web Services. This research was
conducted independently of that employment, did not use proprietary
or employer-confidential data, systems, or resources, and the views
expressed are the authors' own and do not necessarily reflect the
position of Amazon Web Services or Amazon.com, Inc. The authors
declare no other conflicts of interest.

\FloatBarrier

\end{document}